\documentclass[amsmath, superscriptaddress, amssymb, aps, prd,reprint,longbibliography,nofootinbib,twocolumn, notitlepage,10pt]{revtex4-2}
\usepackage{amsmath,amssymb,mathtools}
\usepackage{natbib}
\usepackage{braket}
\usepackage{xcolor}
\RequirePackage[colorlinks,citecolor=blue,urlcolor=magenta,linkcolor=blue]{hyperref}
\usepackage{cleveref}
\usepackage{graphicx}
\usepackage{bm}
\usepackage{subcaption}
\usepackage[normalem]{ulem}
\usepackage{comment}

\renewcommand{\d}{{\rm d}}

\let\calccommentout\iffalse 
\let\calcshow\iftrue

\labelformat{equation}{Eq.~(#1)} 
\labelformat{figure}{Fig.~#1} 
\labelformat{subfigure}{Fig.~\thefigure#1} 
\labelformat{appendix}{Appendix #1}
\begin{document}

\title{Cavity-controlled Inhibition of Decoherence in Accelerated Quantum Detectors}

\author{Harkirat Singh Sahota}
\email{harkirat221@gmail.com}
\affiliation{Centre for Strings, Gravitation and Cosmology, Department of Physics, Indian Institute of Technology Madras, Chennai 600036, India}

\author{Shagun Kaushal}
\email{shagun.kaushal@vit.ac.in}
\affiliation{Department of Physics, School of Advanced Sciences, Vellore Institute of Technology, Vellore 632014, India}

\author{Kinjalk Lochan}
\email{kinjalk@iisermohali.ac.in}
\affiliation{Department of Physical Sciences, Indian Institute of Science Education \& Research (IISER) Mohali, Sector 81 SAS Nagar,
Manauli PO 140306 Punjab India}

\date{\today}

\begin{abstract}

Vacuum fluctuations of quantum fields provide an unavoidable environment for any quantum system coupled to it. We study the interplay between boundary conditions and acceleration in determining decoherence of a two-level Unruh-DeWitt detector coupled to a scalar field in a cylindrical cavity. We show that the decoherence rate closely follows the emission profile, and exhibits {\it Purcell-like} enhancement for both inertial and uniformly accelerated detectors. The acceleration induces an effective smearing of the resonant density of states, diluting the resonance enhancement for large accelerations while replacing the inertial off-resonant decay with an oscillatory behavior for small accelerations. For moderate accelerations, this interplay between cavity-induced and acceleration-assisted effects results in an extended region of cavity parameters where decoherence is strongly suppressed, particularly in regimes where the inertial detector otherwise experience strong decoherence. Thus, contrary to naive expectations, the Unruh thermality in a suitably engineered cavity can enhance rather than degrade quantum coherence, providing a very uncharacteristic feature of quantum fields in non-inertial frames.
\end{abstract}

\maketitle

\section{Introduction}

An intriguing feature of quantum field theory is that the vacuum state, while devoid of particle excitations, nevertheless exhibits ubiquitous fluctuations.
These vacuum fluctuations further play a pivotal role in a wide range of physical phenomena, e.g., Casimir effect, Schwinger effect, Hawking-Unruh effect \cite{Casimir:1948dh,Schwinger:1951nm,Hawking:1974rv,Unruh:1976db}. At the heart of the many of these predictions is the idea that the notion of a vacuum in quantum field theory is observer dependent when one goes beyond inertial systems: different observers may associate different particle content to the same field state \cite{Fulling:1972md,Birrell:1982ix}. An archetypical example that we focus on is the Unruh effect, where a uniformly accelerated observer perceives the inertial vacuum as a thermal bath \cite{Unruh:1976db}. A standard operational framework for probing these observer-dependent vacuum properties is provided by Unruh–DeWitt (UDW) detectors \cite{Unruh:1976db,DeWitt:1980hx}--idealized two-level system locally coupled to a quantum field--which allows one to translate field fluctuations into quantum observables associated with the detector.

Although the conceptual underpinning of the Unruh effect is profound, owing to its parallels with the Hawking effect, the direct observation of Unruh effect in a realistic experimental settings remains a challenge, casting a shadow of uncertainty on this otherwise powerful framework \cite{Crispino:2007eb}. The accelerations required to produce observable transition rates are typically huge, and the resulting detector responses are often exceedingly small at achievable scales \cite{Bell:1982qr,Bell:1986ir,Barut:1990uy,Unruh:1998gq}. These challenges have motivated the exploration of alternative strategies to amplify or control detector–field interactions in experimentally accessible regimes to probe the quantum artifacts of non-inertial motions \cite{Rogers:1988zz,Yablonovitch:1989zza,Raval:1995mb,Chen:1998kp,Vanzella:2001ec,Scully:2003zz,Belyanin:2006,Schutzhold:2008zza,Aspachs:2010hh,Retzker:2007vql}.

One promising approach in this regard is the use of cavities to modify the field mode structure \cite{Scully:2003zz}. By altering the spectral density of vacuum fluctuations, cavities can significantly enhance or suppress detector responses relative to free space, rendering the acceleration-induced effects within the reach of experimental observation \cite{Lochan:2019osm,Stargen:2021vtg,Soda:2022}. In fact, quantum observables like the geometric phase \cite{Martin-Martinez:2010gnz,Arya:2022lay,Barman:2024jpc,Ghosh:2024mqy}, entanglement \cite{Zhou:2023xag,Mukherjee:2023bnt} or the phenomenon like radiative shifts \cite{Arya:2023okh,Arya:2024qke,Peng:2024pfy} or superradiance \cite{Deswal:2025cjw,Zheng:2024drg} has been explored as possible probes to the non-inertial effects in related setups.

Apart from imparting thermal environment to accelerating systems, these vacuum fluctuations also play a central role in ascribing nonlocal quantum correlations (often referred to as “spooky action at a distance”) \cite{Summers:1985pzz,Summers:1987ze} resulting in entanglement harvesting from the vacuum \cite{Reznik:2002fz,Reznik:2003mnx} and provide inescapable environment \cite{Kiefer:1992cn,Ford:1993re} to such systems.
Therefore, they are quite capable of invoking quantum fluctuations-induced decoherence \cite{Zeh:1970zz,Zurek:1982ii} to practically any physical system.  Decoherence is the mechanism which makes an open system shed its quantum coherence through interaction with environment. 
The decoherence rate from such  quantum fluctuations is typically computed  within a regime of approximations such as Markov approximation and  rotating wave approximation, see \cite{Han:2025pql} for detailed discussion. Thus decoherence not only entails environmental details to the system but also provides an avenue to ascertain the inner workings of the very fundamentals of quantum theory in action, see \cite{Deleglise2008-hv,Pabst:2011,Bourassin-Bouchet:2020,Kaplanek:2019dqu,Kaplanek:2019vzj,Kaplanek:2020iay,Burgess:2022nwu,Colas:2024xjy,Burgess:2024heo,Burgess:2025dwm,Christie:2025knc,Gundhi:2024fqv,Gundhi:2025bwj}.

We investigate the decoherence aspects in the energy superpositions of an accelerating two-level atom inside a cylindrical cavity in this work. The vacuum fluctuation driven decoherence timescale of energy superposed state of inertial two-level atom is determined by the coupling strength $\lambda$ of the atom with the vacuum fluctuation and the energy gap $\Omega$ as $T_{dec}\sim\lambda^{-2}\Omega^{-1}$ \cite{Anastopoulos:1999ht,Shresta:2004gic}. Presence of real particles in the environment further enhances such decoherence \cite{Zeh:1970zz, Zurek:1982ii}.
The Unruh effect suggests that an accelerated atom perceives the inertial vacuum environment to be thermally populated \cite{Unruh:1976db}. Thus, a natural corollary is that accelerated atoms would enhance the decoherence rate by a thermal factor  \cite{RevModPhys.75.715,Shresta:2004gic,Han:2025pql,Li:2026pxg}. However, we demonstrate that {\it in a controlled cavity environment the accelerated motion and the resulting thermality can in fact be used to suppress the decoherence rate, which is highly counter-intuitive}.

We establish that the decoherence rate for an Unruh-DeWitt detector in the energy superposition is directly tied to its emission rate. For an accelerated  detector, in a free space, the emission rate is thermal following a Planckian spectrum with the Unruh temperature. 
However, this picture gets strongly modified inside a cavity environment. Due to boundary condition driven constraints on the density of modes and their participation in the emission process both become qualitatively different from the free space case \cite{Stargen:2021vtg}.  For a resonant cavity, the emission rate of an inertial detector gets highly amplified at the near resonance, while it falls steeply for off-resonant frequencies \cite{Kleppner:1981, Gabrielse:1985zz}. On the other hand, the divergent resonant structure for the inertial detector gets regularized with the chirp-like oscillatory features for accelerating detector for small accelerations \cite{Stargen:2021vtg}. In fact, if one tunes the radius of cavity in the neighborhood of the resonance points, the acceleration-induced emission rate could be made even larger than the observable inertial emission rate \cite{Stargen:2021vtg}. Thus, the requirement of large accelerations for the observability of the Unruh effect gets traded for the precision in cavity dimensions. However in the same regime, the acceleration driven chirp behavior also causes the emission rate to fall much below its inertial counterpart, suggesting an accelerate induced inhibition to the emission process, thereby leading to a region of inhibited decoherence. There have been attempts to control thermal decoherence by different mechanism \cite{Agarwal:1999,Agarwal:2001, Montina:2008juk}, {\it yet interestingly such suppression of decoherence is more easily achievable at intermediate accelerations, compared to very low or high accelerations, within standard couplings.} In order to demonstrate these features, we undertake a systematic investigation of decoherence aspects of the cavity-detector setup. \\

\section{Decoherence of a monopole under Markovian evolution}
The setup considered in this work consists of a two-level atom with energy gap $\Omega$ coupled with a real scalar field $\phi$ via monopole coupling inside a long cylindrical cavity of radius $R$ and length $L \gg \Omega^{-1}$. 
To study the decoherence, we consider the atom (a quantum detector) to be initially prepared in a superposition state in the energy eigenbasis,
\begin{align}
    \hat\rho_{\text{atom}}^{(0)}
    =\begin{pmatrix}
        p & \sqrt{p(1-p)} \\[4pt]
        \sqrt{p(1-p)} & 1-p
     \end{pmatrix}=\begin{pmatrix}
        \rho_{gg} & \rho_{eg} \\[4pt]
        \rho_{ge} & \rho_{ee}
     \end{pmatrix}
\end{align}
where $g$ and $e$ corresponds to the ground and excited states and parameter $p\in[0,1]$ characterize the interference.
The atom interacts with the scalar field through the monopole coupling
\begin{equation}
    \hat{H}_{\text{int}}
    =\lambda\,\hat\sigma_x(\tau)\,\hat\phi(x(\tau)),
\end{equation}
where the monopole operator in the energy basis is
\begin{equation}
    \label{sigmaxEB}
    \hat{\sigma}_x(\tau)
    = e^{i\Omega\tau}\,|e\rangle\langle g|
      +e^{-i\Omega\tau}\,|g\rangle\langle e|,
\end{equation}
and $\Omega$ is the atomic energy gap.

We assume the field to be initially in the vacuum state, so that the combined
atom–field density matrix is
\begin{equation}
    \label{rhoeb}
    \hat\rho_{\text{in}}
    =\hat\rho^{(0)}_{\text{atom}}\otimes |0\rangle\langle 0| .
\end{equation}
Being a product state, it does not represent any entanglement initially.
From now on, we shall suppress the tensor product symbol for the sake of notational brevity. The time evolution of this initial state in the interaction picture is given by
\begin{equation}
    \label{outdefR}     \hat{\rho}_\text{out}(\tau)=\hat{U}_{\text{int}}(\tau)\hat{\rho}_\text{in}\hat{U}_{\text{int}}^\dagger(\tau).
\end{equation}
where the interaction picture evolution operator is defined as
\begin{equation}
    \hat{U}_{\text{int}}(\tau)=T ~\exp\bigg(-i\int_0^\tau dt \,\hat{H}_{\text{int}}(t)\bigg)
\end{equation}
$T$ indicates time ordering. On expanding it upto second order in the coupling $\lambda$, and substituting the interaction-picture evolution operator in \eqref{outdefR}, we obtain
\begin{widetext}
    \begin{align}
        \hat\rho_{\text{out}}(\tau)
        =&\, \hat\rho_{\text{in}}-i\int_0^\tau dt\, [\hat{H}_{\mathrm{int}}(t),\hat{\rho}_{\text{in}}]-\!\iint_0^\tau dt dt'\,
        \Theta(t-t')\,
        \big[\hat H_{\mathrm{int}}(t),
            [\hat H_{\mathrm{int}}(t'),\hat{\rho}_{\text{in}}]\big]
        + \mathcal{O}(\lambda^3).
    \end{align}
\end{widetext}
Tracing over the field degrees of freedom gives the reduced atom state,
$
    \hat\rho_{\mathrm{atom}}(\tau)
    = \mathrm{Tr}_{\phi}\!\left[\hat\rho_{\text{out}}(\tau)\right].
$
In the Markovian limit, evolution of the off-diagonal elements of the reduced density matrix of the atom is governed by the master equation \cite{Han:2025pql}
\begin{align}
    \dot{\rho}_{ge}(\tau)=-2\lambda^2\rho_{ge}(\tau)\int_0^\infty\d\tau_-\,\text{Re}\big\{\mathcal{W}(\tau_-)\big\}e^{i\Omega\tau_-}
\end{align}
leading to 
$
    \rho_{eg}(\tau)=e^{-\lambda^2(\mathcal{C}+i\Delta)\tau}\rho_{eg}(0).
$ 
Here the exponents can be identified as the decoherence functional
\begin{align}
    \dot{\mathcal{D}}(\Omega)&=2\int_{0}^{\infty}\!\! \d\tau_-\,\text{Re}\big\{\mathcal{W}(\tau_-)\big\}(\cos{\Omega\tau_-}+i\sin{\Omega\tau_-})\\
    &=\mathcal{C}(\Omega)+i\Delta(\Omega).
      \label{Decoherence}
\end{align}
which govern the evolution of the coherence terms in the reduced density matrix. The real part of the decoherence functional determines the suppression in the off-diagonal terms of the reduced density matrix. Therefore, we term $\mathcal{C}$ {\it the decoherence rate}, which will be the object of our interest in this analysis. 

Another observable that we will discuss throughout is the emission rate of the detector, given by \cite{Crispino:2007eb}
\begin{align}
    \dot{\mathcal{F}}^{em}(\Omega)=\int_{-\infty}^{\infty}\d\tau_-\,\mathcal{W}(\tau_-)e^{i\Omega\tau_-}.
\end{align}

Using the fact that $\mathcal{W}(x_1-x_2)=\mathcal{W}(x_2-x_1)^*\implies\text{Re}\big\{\mathcal{W}(x_1-x_2)\big\}=\text{Re}\big\{\mathcal{W}(x_2-x_1)\big\}$ and $\text{Im}\big\{\mathcal{W}(x_1-x_2)\big\}=-\text{Im}\big\{\mathcal{W}(x_2-x_1)\big\}$, the emission rate can be cast in the form
\begin{equation}
\begin{split}
    \dot{\mathcal{F}}^{em}(\Omega)=\int_{-\infty}^{\infty}\d\tau_-\,\bigg(&\text{Re}\big\{\mathcal{W}(\tau_-)\big\}\cos{\Omega\tau_-}\\
    &-\text{Im}\big\{\mathcal{W}(\tau_-)\big\}\sin\Omega\tau_-\bigg).
\end{split}
\end{equation}
The decoherence rate and transition rates are related by the following relation
\begin{align}
    \mathcal{C}(\Omega)=\frac{1}{2}\left(\dot{\mathcal{F}}^{em}(\Omega)+\dot{\mathcal{F}}^{abs}(\Omega)\right)
\end{align}
where $\dot{\mathcal{F}}^{abs}(\Omega)=\dot{\mathcal{F}}^{em}(-\Omega)$ is the response function giving excitation rate. Thus the decoherence rate and transition rates are directly proportional, with the proportionality constant determined by the detailed balance ratio. In the next subsections, we will investigate the proportionality constant for atom-cavity system.

\subsection{Inertial Decoherence in Cavity}
Using the Wightman function for the scalar field inside a cylindrical cavity
\cite{Stargen:2021vtg},
%
the rest frame decoherence rate  can be written as
\begin{align}
    \mathcal{C}^{\text{In}}&=\frac{2}{(2\pi R)^2}\sum_{m,n}\frac{\text{J}^2_m(\xi_{mn}\rho_0/R)}{\text{J}_{|m|+1}^2(\xi_{mn})}\int_{-\infty}^\infty \frac{\d k_z}{\omega_k}\nonumber\\
    &\qquad\qquad\times\int_0^\infty\d \tau_-\cos{\omega_k\tau_-}\cos{\Omega \tau_-}\\
    &=\frac{1}{2\pi R^2}\sum_{m,n}\frac{\text{J}^2_m(\xi_{mn}\rho_0/R)}{\text{J}_{|m|+1}^2(\xi_{mn})}\nonumber\\
    &\qquad\times\int_{0}^\infty\d\omega_k \frac{\omega_k\Theta(\omega_k-\xi_{mn}/R)}{\sqrt{\omega^2_k-\xi_{mn}^2/R^2}}\underbrace{\frac{\delta(\Omega-\omega_k)}{\omega_k}}_{{\cal I}_0(\Omega,\omega_k)}.
    \label{Cin}
\end{align}
 where $\xi_{mn}$ are the $n-$th zero of the Bessel-J function of order $m$, and we have used the integral representation of Dirac-delta function \cite{gradshteyn2007}. The function ${\cal I}_0(\Omega,\omega_k)$ depicts the participation of the mode functions of the field towards the process of decoherence.

The crucial object that determine the salient features of the cavity physics is the density of field modes, which for fixed order $m$ takes the form \cite{Stargen:2021vtg}
\begin{align}
    \rho(\omega_k)=\sum_{n=1}^\infty\frac{\omega_k/\pi R^2}{\text{J}^2_{|m|+1}(\xi_{mn})}\frac{\Theta(\omega_k-\xi_{mn}/R)}{\sqrt{\omega_k^2-\xi_{mn}^2/R^2}}.
\end{align}
It is diverging when the energy of mode approaches the cavity resonance points $\omega_k\rightarrow\xi_{mn}/R$. 
The integration of the convolution of the two functions over $\omega_k$ leads to
\begin{align}
    \mathcal{C}^{\text{In}}&=\frac{1}{2\pi R}\sum_{m,n}\frac{\text{J}^2_m(\xi_{mn}\rho_0/R)}{\text{J}_{|m|+1}^2(\xi_{mn})}\frac{\Theta(R\Omega-\xi_{mn})}{\sqrt{R^2\Omega^2-\xi_{mn}^2}}\label{Dec_in_fsl}
\end{align}
which turns out to be exactly one half of the emission rate of an inertial detector in cavity $\dot{\tilde{\mathcal{F}}}^{\text{em}}$. Using the free space emission rate $\dot{\tilde{\mathcal{F}}}^{\text{em}}_{\mathcal{M}}=\Omega/2\pi$ and the decoherence rate $\mathcal{C}^{\text{In}}_{\mathcal{M}}=\Omega/4\pi$ (see Appendix \ref{Appendix_fsl}), we arrive at 
\begin{align}
    \frac{\mathcal{C}^{\text{In}}}{\mathcal{C}^{\text{In}}_{\mathcal{M}}}=\frac{\dot{\tilde{\mathcal{F}}}^{\text{em}}}{\dot{\tilde{\mathcal{F}}}^{\text{em}}_{\mathcal{M}}}.
\end{align}
The dimensionless decoherence rate of atom at rest is determined by the ratio of the emission rate in cavity to the emission rate in free space, known as the Purcell enhancement factor \cite{Purcell1946,MabuchiDoherty2002,HarocheRaimond2006}. Since it is well observed  the emission rate of an atom can be selectively amplified inside a cavity, the same holds for the decoherence rate. This amplification occurs as the density of field modes with energy $\omega_k=\xi_{mn}/R$ is strongly enhanced. As a consequence, an atom initially prepared in a superposition state undergoes effectively instantaneous suppression of its quantum coherence, evolving to a mixed state which is now diagonal in the energy eigenbasis, see Appendix \ref{Appendix_Fidelity}. This provides a unique avenue to study decoherence for microscopic systems, where quantum  mechanics is expected to be well verified.


\subsection{Acceleration and Decoherence}
With the insights developed for inertial system in cavity settings, we move on to analyze acceleration induced decoherence. According to the Unruh effect \cite{Unruh:1976db}, an accelerating atom finds itself immersed in a thermal environment, which should enhance the decoherence rate dependent on the temperature it perceives \cite{RevModPhys.75.715,Shresta:2004gic,Han:2025pql,Li:2026pxg}. On the account of change in the participation of modes in the emission process being influenced by the acceleration as discussed in \cite{Stargen:2021vtg}, decoherence rate is expected to be modified as well, in a characteristic fashion distinct to the inertial case.
The Wightman function $\mathcal{W}(\tau_-)$ for the accelerating detector along the trajectory $(a^{-1}\sinh{a\tau},\;\rho_0,\;\theta_0,\;a^{-1}\cosh{a\tau})$ in a cylindrical cavity takes the form
\begin{widetext}
\par\vspace{-\baselineskip}
\begin{equation}
    \mathcal{W}(\tau_-)=\frac{1}{(2\pi R)^2}\sum_{m,n}\frac{\text{J}^2_m(\xi_{mn}\rho_0/R)}{\text{J}_{|m|+1}^2(\xi_{mn})}\!
    \int\! \frac{\d k_z}{\omega_k}e^{-i\frac{\omega_k}{a}(\sinh a\tau-\sinh a\tau')+i\frac{k_z}{a}(\cosh a\tau-\cosh a\tau')}.
\end{equation}
Performing the Lorentz boost transformation on the four-momentum $k_z'=k_z\cosh{a\tau_+}-\omega_k\sinh{a\tau_+}$ and $\omega_k'=\omega_k\cosh{a\tau_+}-k_z\sinh{a\tau_+}$ which preserve the relation $\omega_k'^2=k_z'^2+\xi_{mn}^2/R^2$ and the invariant measure $dk_z/\omega_k$ \cite{Crispino:2007eb}, the Wightman function simplifies to
\begin{equation}
    \mathcal{W}(\tau_-)=\frac{1}{(2\pi R)^2}\sum_{m,n}\frac{\text{J}^2_m(\xi_{mn}\rho_0/R)}{\text{J}_{|m|+1}^2(\xi_{mn})}\!
    \int\! \frac{\d k'_z}{\omega_{k'}}e^{-\frac{2i\omega_{k'}}{a}\sinh{\frac{a\tau_-}{2}}},
\end{equation}
The decoherence rate in this can be expressed as
\begin{align}
    \mathcal{C}
    =&\frac{1}{8\pi^2 R^2}\sum_{m,n}\frac{\text{J}^2_m(\xi_{mn}\rho_0/R)}{\text{J}_{|m|+1}^2(\xi_{mn})}\left[I(\Omega,a)+I(-\Omega,a)+\text{c.c.}\right]\nonumber
\end{align}
where the integral of interest evaluates to \cite{Crispino:2007eb}
\begin{align}
    I(\Omega,a)=\int_{0}^\infty \frac{\d\omega_k}{\sqrt{\omega_k^2-\xi_{mn}^2/R^2}} \omega_k\Theta\left(\omega_k-\frac{\xi_{mn}}{R}\right)\underbrace{\int_{-\infty}^\infty\d \tau_-\frac{e^{i\Omega\tau_-}e^{-2i\omega_k/a\sinh{(a\tau_-/2)}}}{\omega_k}}_{{\cal I}_a(\Omega,  \omega_k)}=\frac{2}{a}e^{-\frac{\pi\Omega}{a}}\text{K}^2_{i\frac{\Omega}{a}}\left(\frac{\xi_{mn}}{aR}\right).
\end{align}
\par\vspace{-\baselineskip}
\end{widetext}
Clearly for the accelerated case the participation function ${\cal I}_a(\Omega,  \omega_k)\sim  \text{K}_{i\frac{2\Omega}{a}}(2\omega_k/a)$ does not only have a single mode support but has a spread out profile in frequency space \cite{Stargen:2021vtg}, owing to the non-linear temporal profile in the Wightman function. The participation function effectively smears the resonant structure of density of states, resulting in regularized decoherence rate.
Using the relation $\text{K}_{i\nu}(x)=\text{K}_{-i\nu}(x),$ decoherence rate for an accelerated detector inside a cylindrical cavity simplifies to
\begin{align}
    \mathcal{C}=&\frac{\cosh{(\pi\Omega/a)}}{\pi^2 R^2a}\sum_{m,n}\frac{\text{J}^2_m(\xi_{mn}\rho_0/R)}{\text{J}_{|m|+1}^2(\xi_{mn})}\text{K}^2_{i\frac{\Omega}{a}}\left(\frac{\xi_{mn}}{aR}\right)\label{Dec_ac_fsl}\\
    &=\frac{1}{2}(1+e^{-2\pi\Omega/a})\dot{\mathcal{F}}^{em}
\end{align}
where $\dot{\mathcal{F}}^{em}$ is the emission rate of the accelerated detector in the cavity \cite{Stargen:2021vtg}. As discussed in the previous section, the proportionality factor is determined from the detailed balance ratio, which is the Gibbs factor indicating the thermal nature of the underlying state even though the response itself is not thermal \cite{Arya:2023okh}, see Appendix \ref{Appendix_Fidelity} for more details. Using the expression of the free-space emission rate for the emission rate \cite{Xu:2023tdt}
\begin{align}
    \dot{\mathcal{F}}^{em}_{\mathcal{M}}=\frac{\Omega}{2\pi}\frac{1}{1-e^{-2\pi\Omega/a}}
\end{align}
we again arrive at the Purcell enhancement in decoherence rate for the accelerating atom in cylindrical cavity;
\begin{align}
    \mathcal{C}=\frac{\Omega}{4\pi}\coth\!\left(\frac{\pi\Omega}{a}\right)\frac{\dot{\mathcal{F}}^{em}}{\dot{\mathcal{F}}^{em}_{\mathcal{M}}}\implies\frac{\mathcal{C}}{\mathcal{C}_\mathcal{M}}=\frac{\dot{\mathcal{F}}^{em}}{\dot{\mathcal{F}}^{em}_{\mathcal{M}}},\label{PEF}
\end{align}
where $\mathcal{C}_\mathcal{M}$ is the thermal decoherence rate of an accelerated detector in free space \cite{Han:2025pql}.
\begin{figure}[t!]
    \centering
    \begin{subfigure}[b]{\columnwidth}
        \includegraphics[width=\linewidth]{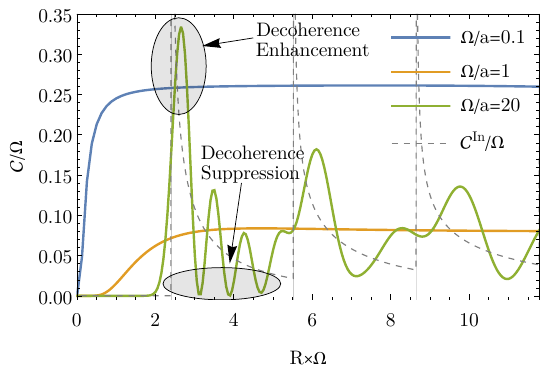}
        \caption{$\mathcal{C}/\Omega$ as a function of the cavity tuning parameter $R\Omega$ for different $\Omega/a$, with dashed curve depicting inertial decoherence rate.}\label{Fig_DR_vs_RORa}
    \end{subfigure}
    \begin{subfigure}[b]{\columnwidth}
        \includegraphics[width=\linewidth]{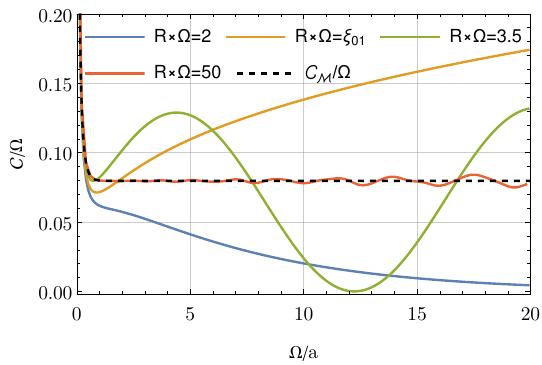}
        \caption{$\mathcal{C}/\Omega$ as a function of $\Omega/a$ for different cavity tuning parameters. The black dashed curve represents the free-space thermal decoherence rate.}\label{Fig_DR_vs_Oa}
    \end{subfigure}
    \caption{Dimensionless decoherence rate $\mathcal{C}/\Omega$ of a uniformly accelerated atom inside a cylindrical cavity. 
   }\label{Fig1}
\end{figure}
\ref{Fig_DR_vs_RORa} displays the decoherence rate $\mathcal{C}/\Omega$ of a uniformly accelerated atom confined within a cylindrical cavity as a function of the cavity tuning parameter $R\Omega$. For very large accelerations the associated length scales of participating modes becomes very small as ${\cal I}_a(\Omega,  \omega_k) \sim  \text{K}_{i\frac{\Omega}{a}}(\omega_k/a)$, is highly suppressed for small $\omega_k$,  the cavity geometry becomes irrelevant and one gets a very good approximation to the free space thermal decoherence. Only smaller  cavity geometries, comparing to the short acceleration scales are able to enforce boundary effects. Consequently the decoherence rate varies smoothly and increases monotonically with the cavity size, rapidly saturating at the inertial free-space thermal decoherence rate~\cite{Han:2025pql}. 

The low acceleration regime, on the other hand, apart from making the cavity effects visible also modulates the participation function ${\cal I}_a(\Omega,  \omega_k)$ in  favor of large wavelength modes. Gradually the Bessel function's oscillatory features for smaller arguments develop and for very low accelerations, the expression \ref{PEF} demonstrates rapid oscillations in the decoherence near the inertial resonant configurations and envelop the monotonic off resonant behavior of the inertial decoherence profile for extremely low accelerations.

{At moderate acceleration scales, it can be seen that the decoherence rate is larger than the inertial and thermal decoherence rate just near the resonance points but more importantly, it gets rapidly suppressed below the inertial as well as thermal baseline if the cavity is judiciously tuned into the decoherence suppressed valleys between the first two resonance points \ref{Fig_DR_vs_RORa}. In fact, both the transition rates and decoherence rates vanish for such a finely tuned cavity, as the atom-field system gets effectively decoupled and the atom undergoes unitary evolution retaining its coherence, see Appendix \ref{Appendix_Fidelity} for behavior of Purity and Fidelity of the atomic state with respect to initial state.} {\it This is highly non trivial and at odds with the expectation, as acceleration induced thermality is typically expected to degrade quantum coherence} \cite{Shresta:2004gic,Han:2025pql,Li:2026pxg}. The acceleration-regulated participation function develops regions of low influence (despite a reasonable  number of available modes) to nullify the effect of enhanced thermal noise. Clearly when the acceleration becomes too high, the participation function also gets dominated with high frequency modes, deficient in being influenced by boundary effects and the Unruh thermality-enhanced decoherence sets in. \ref{Fig_DR_vs_Oa}  provides a consistent picture of how cavity geometry and acceleration jointly control decoherence rate in cavity-atom systems.

\begin{figure}[t]
    \centering
    \includegraphics[width=\linewidth]{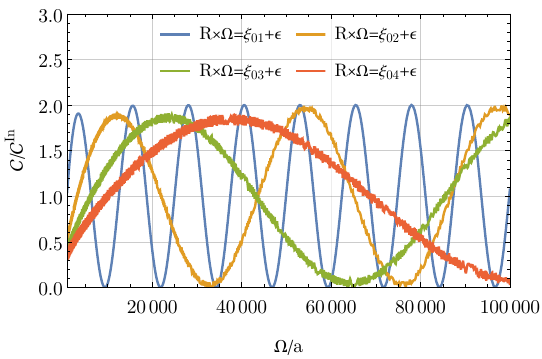}
    \caption{
    Near resonance behavior of the decoherence rate for $\epsilon=10^{-2}$ as a function of dimensionless energy gap near different resonance points.}
    \label{Fig_DRAsy}
\end{figure}

The near resonance character of the decoherence rate, relative to the inertial case, as a function of dimensionless energy gap is shown in the \ref{Fig_DRAsy} for first four resonance configurations. The oscillatory behavior of the decoherence rate  varies for different resonance points. Notably, the amplitude of such oscillations near the resonance configurations are very similar across all the cases considered. 
For higher resonance points, suggesting a larger cavity dimension, the system admits a wider parameter window over which the enhancement or suppression can be fine-tuned, as the region of extrema persist over a larger range of $\Omega/a$. Overall, our investigation reveal a tunable interplay between acceleration and cavity structure that enables controlled modulation of decoherence.  Therefore with moderate acceleration scales $\Omega/a\sim 1-10$ one can construct cavities even with large size and preserve quantum coherence to a very good order. Notably in such geometries, the decoherence rate for an accelerated atom is much smaller than a corresponding inertial atom. It suggests that an accelerated transport may be a more efficient strategy for protecting a quantum state over large distances, compared to an inertial transport.
\section{Discussion}
In this work, we investigated the quantum fluctuations-induced decoherence of an Unruh-DeWitt detector coupled to a scalar field in a cylindrical cavity. Owing to the cavity-modified spectral properties, the response of the detector gets enhanced when the cavity is tuned to the detector energy gap, known as Purcell enhancement \cite{Purcell1946}. While cavity-induced Purcell enhancement is traditionally associated with the emission rate, we show that the decoherence is also determined by the Purcell factor, resulting in loss of coherence of an initial interference state of an inertial detector at specific cavity resonance points.

The acceleration-induced spectral broadening regularize the divergences in density of field modes, leading to finite emission and decoherence rates. For large accelerations (as compared to energy gap scale), the thermal broadening dominates the cavity-induced spectral properties, rendering the decoherence rate effectively as that of free space thermal and insensitive to the cavity resonances. As the acceleration is decreased, resonant behavior re-emerges, revealing the constructive and destructive interplay between thermal broadening and cavity-induced enhancement. This interplay manifests as the oscillatory behavior of the decoherence rate around thermal behavior. The cavity effects for the accelerated detector are most pronounced when the acceleration is extremely small and the decoherence exhibits a chirp-like behavior which envelops the inertial fall-off, same as the behavior observed for emission rate in~\cite{Stargen:2021vtg}.

However, the most striking regime is the one in which the decoherence rate becomes vanishingly small. The detector-cavity system can be tuned such that it exhibit effective decoupling, leading to a strong suppression of both the emission rate and the decoherence rate for an accelerated detector. Interestingly, the inertial detector, in this regime, is still strongly coupled with finite emission and decoherence rates, providing a promising probe of acceleration induced effect. Notably, the tuning is more accessible for moderate acceleration scales (rather than extremely high or low accelerations), where the near-resonant region of decoherence dip persists over a broader range of energy gap and accelerations, allowing achievable control over the suppression of decoherence. In this regime, the environment-induced loss of coherence is strongly suppressed, effectively freezing the dissipative dynamics and stabilizing the quantum state over longer timescales, in a sharp contrast to the expectations from thermality-enhanced decoherence. Thus, acceleration with a fine-tuned cavity can aid rather than hinder the quantum coherence. This provides a unique probe of QFT in non-inertial frames and also opens up avenues of studies for acceleration assisted processes of low environmental disturbances where the quantum fluctuations driven processes can be further suppressed by tuning both the cavity and the accelerated motion. Further, cavity guided transportation of atomic systems on accelerating trajectories may provide a robust way of decoherence controlled transportation of initial state over long length scales, which will also ease away the requirement of extremely precise cavities or highly controlled accelerations, through the tolerance windows they offer unlike for inertial systems.


\paragraph*{Acknowledgments\textemdash}
Research of K.L. is partially supported by ANRF, Government of India, through a  research grant no. CRG/2023/004641.  S.K. acknowledges Vellore Institute of Technology (VIT), Vellore, for financial support provided through the Seed Grant for this research.


\appendix

\section{Free-space limit}\label{Appendix_fsl}
In this appendix, we show the free-space limit of the decoherence rate $R\rightarrow\infty$ reproduce vacuum and thermal decoherence for inertial and accelerated detectors. Introducing $k_\perp=\xi_{mn}/R$ and using $\sum_n\to (R/\pi)\int dk_\perp$ together with $\text{J}_{|m|+1}^2(\xi_{mn})\simeq 2/(\pi R k_\perp)$, decoherence rate for inertial detector in \ref{Dec_in_fsl}
\begin{equation}
\mathcal{C}^{\text{In}}=\sum_m \int_0^\infty \frac{k_\perp\,dk_\perp}{4\pi}
\text{J}_m^2(k_\perp\rho_0)
\frac{\Theta(\Omega-k_\perp)}{\sqrt{\Omega^2-k_\perp^2}}.
\end{equation}
Using Neumann's addition theorem \cite{NIST:DLMF} $\sum_m \text{J}_m^2(k_\perp\rho_0)=1$, the free-space limit evaluates to
\begin{equation}
\mathcal{C}^{\text{In}}_{\mathcal{M}}=\int_0^{\Omega}\frac{k_\perp\,dk_\perp}{4\pi\sqrt{\Omega^2-k_\perp^2}}=\frac{\Omega}{4\pi}.
\end{equation}
which is the ambient decoherence due to the vacuum fluctuations \cite{Han:2025pql}.

Similarly for accelerated detector, the free-space limit of decoherence rate in \ref{Dec_ac_fsl} evaluates to
\begin{align}
    \mathcal{C}=&\frac{\cosh{(\pi\Omega/a)}}{\pi^2 R^2a}\frac{\pi R}{2}\frac{R}{\pi}\int_0^\infty k_\perp\d k_\perp \text{K}^2_{i\frac{\Omega}{a}}\left(\frac{k_\perp}{a}\right)\nonumber\\
    =&\frac{\cosh{(\pi\Omega/a)}}{2\pi^2 a}\frac{\Omega a\pi}{2\sinh(\pi\Omega/a)}=\frac{\Omega}{4\pi}\coth(\pi\Omega/a)
\end{align}
which is the thermal decoherence rate with the temperature of the bath being $a/2\pi$ \cite{Han:2025pql}.

\section{Resonant decoherence for detector with small acceleration}\label{Appendix_res}
\begin{figure}
    \centering
    \includegraphics[width=\linewidth]{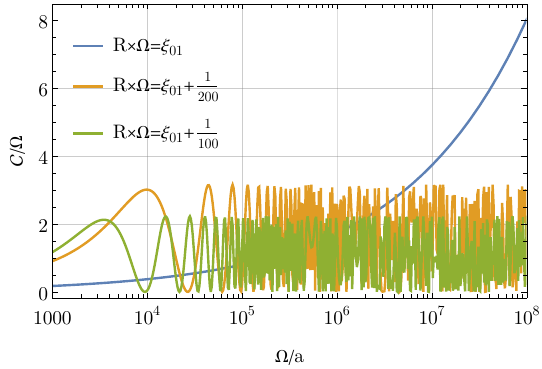}
    \caption{At resonance versus near resonance behavior of dimensionless decoherence rate as a function of dimensionless energy gap at first resonance point. }    \label{Fig_DRAsy_App}
\end{figure}
The near- and at resonant behavior of the decoherence rate for small acceleration is shown in \ref{Fig_DRAsy_App}. The divergent decoherence rate at the resonance points is regularized due to thermal broadening for the accelerated detector leading to oscillatory behavior with decaying amplitude (a chirp-like behavior) in between the resonance points, just like for emission rate of the accelerated detector in cylindrical cavity reported in \cite{Stargen:2021vtg}. 
We see for a relatively larger acceleration, the near resonance decoherence rate is large as compared to its at resonance value. As we keep on decreasing the acceleration, the near resonance decoherence rate keeps on oscillating with the same amplitude. On the other hand, owing to its divergent behavior for the inertial atom, the decoherence rate at resonance point keeps on increasing monotonically and supersede the near resonance enhancement for smaller accelerations.

\section{Fidelity and purity of a monopole under Markovian evolution}\label{Appendix_Fidelity}
The time evolution of the reduced density matrix of a monopole atom coupled to a scalar field, in the Markov approximation, is given by \cite{Han:2025pql}
\begin{widetext}
\begin{align}
    \dot{\rho}_{gg}(t)&=\lambda^2\left(\int_{-\infty}^\infty\d\tau~e^{-i\Omega\tau}\mathcal{W}(\tau)-2\rho_{gg}(t)\int_0^\infty\d\tau~\text{Re}\{\mathcal{W}(\tau)\}\cos\Omega\tau\right)=\lambda^2(\dot{\mathcal{F}}^{abs}(\Omega)-2\rho_{gg}(t)\mathcal{C}(\Omega))\\
    \dot{\rho}_{ee}(t)&=\lambda^2\left(\int_{-\infty}^\infty\d\tau~e^{i\Omega\tau}\mathcal{W}(\tau)-2\rho_{ee}(t)\int_0^\infty\d\tau~\text{Re}\{\mathcal{W}(\tau)\}\cos\Omega\tau\right)=\lambda^2(\dot{\mathcal{F}}^{em}(\Omega)-2\rho_{ee}(t)\mathcal{C}(\Omega))\\
    \dot{\rho}_{ge}(t)&=-2\lambda^2\left(\int_0^\infty\d\tau~\text{Re}\{\mathcal{W}(\tau)\}\cos\Omega\tau+i\int_0^\infty\d\tau~\text{Re}\{\mathcal{W}(\tau)\}\sin\Omega\tau\right)\rho_{ge}(t)=-\lambda^2(\mathcal{C}(\Omega)+i\Delta(\Omega))\rho_{ge}(t).
\end{align}
The solution of these equation takes the form
\begin{align}
    \!\!\!\rho_{gg}(t)&=e^{-2\lambda^2\mathcal{C}(\Omega)t}\left(\!\rho_{gg}(0)-\frac{\dot{\mathcal{F}}^{abs}(\Omega)}{2\mathcal{C}(\Omega)}\!\right)+\frac{\dot{\mathcal{F}}^{abs}(\Omega)}{2\mathcal{C}(\Omega)}\\
    \rho_{ee}(t)&=e^{-2\lambda^2\mathcal{C}(\Omega)t}\left(\rho_{ee}(0)-\frac{\dot{\mathcal{F}}^{em}(\Omega)}{2\mathcal{C}(\Omega)}\right)+\frac{\dot{\mathcal{F}}^{em}(\Omega)}{2\mathcal{C}(\Omega)}\\
    \rho_{ge}(t)&=\rho_{ge}(0)e^{-\lambda^2(\mathcal{C}(\Omega)+i\Delta(\Omega))t},
\end{align}
where $\rho_{gg}(0)=p=1-\rho_{ee}(0)$ and $\rho_{ge}(0)=\sqrt{p(1-p)}$ for the interference state under consideration. Here we see, when $\mathcal{C}(\Omega)=0$ and but the ratio of response and decoherence rate is finite (as is the case for the atom in cavity), the system evolves unitarily as $\rho_{gg/ee}(t)=\rho_{gg,ee}(0)$ and $\rho_{ge}(t)=|\rho_{ge}(0)|$.

In the late-time ($t\gg\lambda^{-2}\mathcal{C}^{-1}$) limit, the state evolves to a mixed state where the off diagonal elements of reduced density matrix vanish while the diagonal elements are determined by the ratio of response and decoherence rate as
\begin{align}
    \lim_{t\rightarrow\infty}\rho_{gg}(t)=\frac{\dot{\mathcal{F}}^{abs}(\Omega)}{2\mathcal{C}(\Omega)},\qquad
    \lim_{t\rightarrow\infty}\rho_{ee}(t)=\frac{\dot{\mathcal{F}}^{em}(\Omega)}{2\mathcal{C}(\Omega)},\qquad
    \lim_{t\rightarrow\infty}\rho_{ge}(t)=0=\lim_{t\rightarrow\infty}\rho_{eg}(t)
\end{align}
For an accelerated detector in the free space, the response function and decoherence rate takes the form \cite{Xu:2023tdt}
\begin{align}
    \dot{\mathcal{F}}^{em}(\Omega)=\frac{\Omega}{2\pi}\frac{1}{e^{2\pi\Omega/a}-1},\qquad
    \mathcal{C}(\Omega)=\frac{\Omega}{4\pi}\coth\left(\frac{\pi\Omega}{a}\right),
\end{align}
while the imaginary part of the decoherence functional $\Delta(\Omega)$ has UV divergences \cite{Han:2025pql}, which are expected to be regularized when the finite extent of the detector is considered \cite{Schlicht:2003iy,Louko:2006zv}. The asymptotic late-time $t\rightarrow\infty$ state of the accelerated detector is the Gibbs state \cite{Han:2025pql}
\begin{align}
    \!\!\!\!\!\!\rho_{gg}(t)\bigg|_{t\rightarrow\infty}\!\!\!\!\!\!=\frac{1}{e^{2\pi\Omega/a}+1},\qquad\rho_{ee}(t)\bigg|_{t\rightarrow\infty}\!\!\!\!\!\!=\frac{1}{e^{-2\pi\Omega/a}+1}
\end{align}
On the other hand, for cavity-atom system we have
\begin{align}
    \mathcal{C}(\Omega)&=\frac{1}{2}(1+e^{-2\pi\Omega/a})\dot{\mathcal{F}}^{em}(\Omega)=\frac{1}{2}(1+e^{2\pi\Omega/a})\dot{\mathcal{F}}^{abs}(\Omega)\label{CWRel1}
\end{align}
as decoherence rate in \ref{Dec_ac_fsl} is an even function of energy gap. Therefore, the density matrix of an accelerator detector in cavity does evolve to the Gibbs state in the asymptotic future. In this sense, the cavity-modified vacuum acts as an effective thermal bath for the accelerated detector, despite the detector response not being equal to the response of inertial detector in a thermal state in cavity. This provides a stronger characterization of thermality than that inferred solely from detailed balance of transition rates. While it has been argued that accelerated atoms inside a cavity do not perceive vacuum fluctuations as a thermal bath \cite{Hu-Roura:2004,Scully:2004,Hu:2004zu}, our analysis shows that the detector nevertheless thermalizes at the level of its reduced state. This is consistent with earlier observations on the thermal nature of cavity-modified vacuum for accelerated observer \cite{Belyanin:2006,Arya:2024qke}.

We are interested in the decoherence free subspace of the atom-cavity setup, which can be identified by looking at the fidelity of the detector state $\rho(t)$ compared to the initial state $\rho_0$ is
\begin{align}
    F(\rho,\rho_0)&=\text{Tr}\{\sqrt{\sqrt{\rho}\rho_0\sqrt\rho}\}^2\xrightarrow{\text{2-dim H.S.}}\text{Tr}\{\rho\rho_0\}+\sqrt{\text{det}(\rho)\text{det}(\rho_0)}\xrightarrow{\rho\text{ or }\rho_0\text{ pure state}}\text{Tr}\{\rho\rho_0\}\\
    &=\rho_{gg}(0)\rho_{gg}(t)+\rho_{ee}(0)\rho_{ee}(t)+\rho_{ge}(0)(\rho_{ge}(t)+\rho^*_{ge}(t))\\
    &=p\frac{\dot{\mathcal{F}}^{abs}(\Omega)}{2\mathcal{C}(\Omega)}+(1-p)\frac{\dot{\mathcal{F}}^{em}(\Omega)}{2\mathcal{C}(\Omega)}+e^{-2\lambda^2\mathcal{C}(\Omega)t}\biggr(-p\frac{\dot{\mathcal{F}}^{abs}(\Omega)}{2\mathcal{C}(\Omega)}-(1-p)\frac{\dot{\mathcal{F}}^{em}(\Omega)}{2\mathcal{C}(\Omega)}+p^2+(1-p)^2\biggr)\nonumber\\
    &~~~~~~~~~~+2p(1-p)e^{-\lambda^2\mathcal{C}(\Omega)t}\cos(\Delta(\Omega)t)
\end{align}
where the atom is prepared in an initially pure state. For the cavity-atom system, we get
\begin{align}
    F(\rho,\rho_0)&=\frac{p}{1+e^{2\pi\Omega/a}}+\frac{1-p}{1+e^{-2\pi\Omega/a}}+e^{-2\lambda^2\mathcal{C}(\Omega)t}\biggr(-\frac{p}{1+e^{2\pi\Omega/a}}-\frac{1-p}{1+e^{-2\pi\Omega/a}}+p^2+(1-p)^2\biggr)\nonumber\\
    &~~~~~~~~~~+2p(1-p)e^{-\lambda^2\mathcal{C}(\Omega)t}\cos(\Delta(\Omega)t)
\end{align}
In the asymptotic $t\rightarrow\infty$ case, fidelity takes the form
\begin{align}
    F(\rho,\rho_0)&=\frac{1}{2}\left(1+(1-2p)\tanh{\frac{\pi\Omega}{a}}\right).
\end{align}
However, in the decoherence-free regime $\mathcal{C}=0$, we get
\begin{align}
    F(\rho,\rho_0)&=1-2p(1-p)(1-\cos(\Delta(\Omega)t)).
\end{align}
fidelity is determined by the UV-divergent $\Delta(\Omega)$, which appears as unitary phase evolution and does not indicate dissipative decoherence.  

Another useful measure is the purity of the final state $\rho$
\begin{align}
    \mathcal{P}(\rho)&=\text{Tr}\{\rho^2\}=\rho_{gg}(t)^2+\rho_{ee}(t)^2+2\rho_{ge}(t)^2\\
    &=\frac{\dot{\mathcal{F}}^{abs}(\Omega)^2+\dot{\mathcal{F}}^{em}(\Omega)^2}{4\mathcal{C}(\Omega)^2}+e^{-2\lambda^2\mathcal{C}(\Omega)t}\left(\frac{\dot{\mathcal{F}}^{em}(\Omega)}{\mathcal{C}(\Omega)}-\frac{\dot{\mathcal{F}}^{abs}(\Omega)^2+\dot{\mathcal{F}}^{em}(\Omega)^2}{2\mathcal{C}(\Omega)^2}+p\left(2+\frac{\dot{\mathcal{F}}^{abs}(\Omega)-\dot{\mathcal{F}}^{em}(\Omega)}{\mathcal{C}(\Omega)}\right)-2p^2\right)\nonumber\\
    &\qquad +e^{-4\lambda^2\mathcal{C}(\Omega)t}\left(1-\frac{\dot{\mathcal{F}}^{em}(\Omega)}{\mathcal{C}(\Omega)}+\frac{\dot{\mathcal{F}}^{abs}(\Omega)^2+\dot{\mathcal{F}}^{em}(\Omega)^2}{4\mathcal{C}(\Omega)^2}-p\left(2+\frac{\dot{\mathcal{F}}^{abs}(\Omega)-\dot{\mathcal{F}}^{em}(\Omega)}{\mathcal{C}(\Omega)}\right)+2p^2\right)
\end{align}
using the relations in \ref{CWRel1}, we get
\begin{align}
    \mathcal{P}(\rho)&=\frac{1+\tanh^2{\pi\Omega/a}}{2}+e^{-2\lambda^2\mathcal{C}(\Omega)t}\left(2p-2p^2+\tanh{\pi\Omega/a}-2p\tanh{\pi\Omega/a}-\tanh^2{\pi\Omega/a}\right)\nonumber\\
    &\qquad+\frac{1}{2}e^{-4\lambda^2\mathcal{C}(\Omega)t}(-1+2p+\tanh{\pi\Omega/a})^2
\end{align}
The two limits of interest are the asymptotic case $t\rightarrow\infty$
\begin{align}
    \mathcal{P}(\rho)&=\frac{1+\tanh^2{\pi\Omega/a}}{2}
\end{align}
corresponding to the purity of a thermal state as expected. This limit also matches to the case when the decoherence rate diverge, for example at the resonance points. For the case when the cavity-atom system is tuned to decoherence-free regime $\mathcal{C}(\Omega)\rightarrow0$, we get
\begin{align}
    \mathcal{P}(\rho)=1.
\end{align}
Thus, the state remains pure in the decoherence-free regime.


\end{widetext}
%

\end{document}